\documentclass[a4paper, UTF8, 12pt, fleqn]{article}
\usepackage[left=1.3cm, right=1.3cm, top=1.3cm, bottom=2cm]{geometry}
\usepackage{amsmath,graphicx,subfig,color}

\usepackage{indentfirst}
\usepackage{cite}
\usepackage[colorlinks,
linkcolor=blue,
anchorcolor=blue,
citecolor=blue
]{hyperref}

\title{A new approach for the inversion of residual stress based on acoustoelasticity theory and full waveform inversion}
\author{Maoyu Xu\textsuperscript{1}, Hongjian Zhao\textsuperscript{1}, Changsheng Liu\textsuperscript{2}, Yu Zhan\textsuperscript{1}$^,$\thanks{Corresponding author: zhanyu@mail.neu.edu.cn}}

\date{}
\begin{document}
\footnotetext[1]{College of Sciences, Northeastern University, Shenyang, 110819, China}  
\footnotetext[2]{Key Laboratory for Anisotropy and Texture of Materials Ministry of Education, Northeastern University, Shenyang, 110819, China}  
\maketitle
\section*{Abstract}
Acoustoelasticity theory has been widely used to evaluate the residual stress (or prestress), almost all the available ultrasonic stress detection methods are based on the relationship between the magnitude of stress and wave speed, but these measurement methods make the assumption that the stress is uniform, only one point or average stress in the direction of ultrasound propagation can be obtained. However, the real stress distribution is usually nonuniform. In order to obtain the stress distribution in the direction of ultrasound propagation, in this paper, we propose a new approach: the inversion of residual stress. In the theory part, the inversion of residual stress is transformed into an optimization problem. The objective function is established, and the gradient of the objective function to the stress is derived using the adjoint method, which has been maturely applied in full waveform inversion. In the numerical simulation part, the welding process is simulated using the finite element method to obtain a database of the residual stress field. Then the residual stress is evaluated by inversion approach and the influence of the number of sources and receivers and the frequency of the excitation wave on the inversion effect is discussed. The results show that the inversion of residual stress is still challenging with a small amount of data, but a more accurate inversion can be obtained by appropriately increasing the number of sources and receivers. This study provides an appropriate method for the evaluation of residual stress distribution and lays the theoretical and simulation foundation for the application of ultrasonic stress testing in it.

\vspace{2em} 
\noindent \textbf{Keywords:} Residual stress distribution; Acoustoelasticity; Inverse problem; Full waveform inversion;

\maketitle

\section{Introduction}

With the development of acoustoelastic theory \cite{PhysRev.92.1145, thurston1964third}, ultrasonic stress detection technology has been widely applied due to its advantages of non-destructive, low cost, and high accuracy. Pao \cite{Pao1985AcoustoelasticWI} derived the theory of acoustoelasticity, studied the propagation of ultrasound in orthotropic elastic solids with initial stress, and derived the relationship between strain and wave velocity when the coordinate system coincides with the principal strain axis. Duquennoy \cite{Duquennoy1999InfluenceON} provided a more detailed introduction to residual stress evaluation from theory and experiment. Acoustoelastic theory has been widely applied in loading stress analysis, residual stress assessment, and geotechnical engineering \cite{HUANG2023102832, LIU2022108603,WANG2022108500}. However, most of the current residual evaluation methods are based on the relationship between the stress and the wave velocity magnitude, which can only obtain the stress at a point or the average stress within the detection distance. Since the actual prestress is usually nonuniformly distributed, if the detection distance is too long, the error will increase. A novel approach to stress evaluation is proposed. Inspired by full waveform inversion (FWI), this method utilizes FWI concepts and tools to address the challenge of characterizing nonuniform residual stress.

The proposed residual stress inversion approach is fundamentally framed as an inverse problem. In mechanical problems, the relationship between input parameters and output variables is often represented by \( g(x,p) = 0 \), where \( p \) defines inputs like boundary conditions or material properties, and \( x \) represents outputs such as displacement or stress at specific points. A forward problem involves determining the output \( x \) from known parameters \( p \). Conversely, an inverse problem seeks to infer the parameters \( p \) from observed outputs \( x \) \cite{CAO2002171}.

FWI is a well-established inverse problem in geophysics. It employs acoustic, elastic, or anelastic wave equations to achieve high-accuracy seismic imaging of Earth's interior structure \cite{2022-031}. In the FWI, $g(x,p) = 0$ is the wave equation, $p$ is the wave velocity varying with spatial position, and $x$ is the displacement of some points. FWI achieves high-resolution seismic imaging by iteratively minimizing the misfit between observed and simulated waveforms, enabling inversion of subsurface properties such as velocity and density to reconstruct complex structures \cite{nolet1987seismic}. J. Virieux \cite{10.1190/1.3238367} provides a detailed overview to the development history and main processes of FWI. Jeroen Tromp \cite{10.1111/j.1365-246X.2004.02453.x} introduces the concept of time reversal and banana-doughnut kernels from an abstract theoretical perspective. With the improvement of FWI, some mature theories and tools can be used for nonuniform residual stress inversion, such as adjoint method. In the inversion theory section, we describe how these methods can be applied to the new residual stress evaluation approach. Extensive FWI experiments also offer a foundation of reliability for future experimental investigations into residual stress inversion.

In this paper, we will directly use the relationship between displacement and stress to evaluate stress, rather than the magnitude of wave velocity or travel time. This study is conducted entirely through numerical simulations. In the theory section, firstly, we will briefly introduce the theory of acoustoelasticity and the equation between displacement and stress in the case of uniaxial compression. The theoretical derivation of stress inversion is presented next. In this section, the use of time reversal and the adjoint method is detailed to compute the gradient of the objective function with respect to stress. The proposed inversion flowchart provides a straight forward framework for residual stress inversion. In the numerical simulation part, the database of residual stress is obtained by simulating the welding process using the finite element method, and the residual stress inversion is performed according to the theory derived above. The impact of the number of excitation and receiving points on inversion accuracy is analyzed. The results indicate that with limited data, the inversion performs poorly. However, as the number of excitation and receiving points increases, the results progressively approximate the true stress distribution with high accuracy. And stress inversion remains challenging when the frequency of the excitation wave is high. Based on the above study, the correctness of the inverse theory derivation is proved and the high accuracy of the new residual stress evaluation method is demonstrated. The proposed NDT method holds significant potential for achieving high-precision detection of residual stress distributions. 

\section{Theory}
\subsection{Acoustoelasticity theory}
For the theory of acoustoelasticity, three states of the medium are adopted: the configuration without stress and strain is called the nature state and is represented by the coordinates $\xi_{\alpha}(\alpha=1,2,3)$; the configuration under static deformation is called the initial state and is represented by the coordinates $X_{J}(J=1,2,3)$; the configuration of initial state superimposed with small acoustic perturbations is called the final state and is represented by the coordinates $x_{j}(j=1,2,3)$. Pao derived the equations of acoustoelasticity in natural frame of reference \cite{Pao1985AcoustoelasticWI}. For homogeneous predeformed medium: 
\begin{equation}\label{eq:govering}
	\frac{\partial}{\partial\xi_\beta}P^n_{\alpha\beta}=\rho^o\frac{\partial^2u_\alpha}{\partial t^2},
\end{equation}
where
\begin{equation}
	P^n_{\alpha\beta} = A^{n}_{\alpha\beta\gamma\delta}\frac{\partial u_\gamma}{\partial\xi_\delta},
\end{equation}
\begin{equation}
	A^{n}_{\alpha\beta\gamma\delta}=s_{\beta\delta}^i\delta_{\alpha\gamma}+\Gamma_{\alpha\beta\gamma\delta}=C_{\beta\delta\zeta \eta}\varepsilon^{i}_{\zeta \eta }\delta_{\alpha\gamma}+\Gamma_{\alpha\beta\gamma\delta},
\end{equation}
\begin{equation}
	\Gamma_{\alpha\beta\gamma\delta}=C_{\alpha\beta\gamma\delta}+C_{\alpha\beta\lambda\delta}\frac{\partial u_{\gamma}^{i}}{\partial\xi_{\lambda}}+C_{\lambda\beta\gamma\delta}\frac{\partial u_{\alpha}^{i}}{\partial\xi_{\lambda}}+C_{\alpha\beta\lambda\delta\zeta\eta} \varepsilon^{i}_{\zeta\eta}.
\end{equation}

The above formula is valid under the condition of small predeformation.
\begin{equation}\label{eq:small_dis1}
	s_{\beta\delta}^i = C_{\beta\delta\zeta \eta}\varepsilon^{i}_{\zeta \eta },
\end{equation}
\begin{equation}\label{eq:small_dis2}
	\varepsilon^{i}_{\alpha\beta}=\frac12\left(\frac{\partial u_\alpha^i}{\partial\xi_\beta}+\frac{\partial u_\beta^i}{\partial\xi_\alpha}\right).
\end{equation}
where the variables $P^n_{\alpha\beta}$, $A^n_{\alpha\beta\gamma\delta}$ and $\Gamma_{\alpha\beta\gamma\delta}$ have no physical significance, they are introduced solely to simplify the equations; $C_{\alpha\beta\gamma\delta}$ and $C_{\alpha\beta\lambda\delta\zeta\eta}$ are the second and third order elastic constants; $u_\alpha^i$ is the predeformed displacement from the natural state to the initial state; $\varepsilon_{\alpha\beta}^{i}$ represents the initial Cauchy strain tensor; $s_{\alpha\beta}^i$ is the second Piola-Kirchhoff (P–K) prestress tensor in the nature configuration; $u_\alpha$ is the incremental displacement from the initial state to the finial state; $\rho^o$ is the density in nature configuration. 

The time-harmonic plane wave is assumed
\begin{equation}\label{eq:harmonic}
	u_\alpha=U_\alpha\exp[i(KN_\lambda \xi_\lambda-\omega t)].
\end{equation}
where $U_{\alpha}$ is the amplitude, $K$ is the wave number, $N_{\lambda}$ is the component of the wave normal, $\omega$ is the angular frequency, and the wave velocity $ v = \omega / k$. 

If the plane wave propagating in the direction of the $\xi_3$ axis, substituting Eq.~(\ref{eq:harmonic}) into Eq.~(\ref{eq:govering})
\begin{equation}\label{eq:4}
	(A^n_{\alpha3\gamma3}-\rho^o v^2\delta_{\alpha\gamma})U_\gamma=0,
\end{equation}
Eq.~(\ref{eq:4}) has a non-zero solution if the coefficient determinant is zero \cite{10.3389/feart.2022.886920},
\begin{equation}
	\det\Big[A^n_{\alpha3\gamma3}-\rho^o \nu^{2}\delta_{\alpha\gamma}\Big]=0,
\end{equation}

Voigt's contracted notation is adopted \cite{Pao1985AcoustoelasticWI}
\begin{equation}
	\left[\begin{array}{ccc}
		A_{55}-\rho^{o} \nu^{2} & A_{54} & A_{53} \\
		A_{45} & A_{44}-\rho^{o} \nu^{2} & A_{43} \\
		A_{35} & A_{34} & A_{33}-\rho^{o} \nu^{2}
	\end{array}\right]=0,
\end{equation}
where
\begin{equation}
	\begin{aligned}
		&A_{33}=T_{33}^{i}+c_{33}\left(1+2 \varepsilon_{33}^{i}\right)+c_{331} \varepsilon_{11}^{i}+c_{332} \varepsilon_{22}^{i}+c_{333} \varepsilon_{33}^{i}, \\
		&A_{44}=T_{44}^{i}+c_{44}\left(1+2 \varepsilon_{22}^{i}\right)+c_{441} \varepsilon_{11}^{i}+c_{442} e_{22}^{i}+c_{443} \varepsilon_{33}^{i}, \\
		&A_{55}=T_{55}^{i}+c_{55}\left(1+2 \varepsilon_{11}^{i}\right)+c_{551} \varepsilon_{11}^{i}+c_{552} \varepsilon_{22}^{i}+c_{553} \varepsilon_{33}^{i}, \\
		&A_{34}=A_{43}=c_{33}\left(\partial u_{2}^{i} / \partial \xi_{3}\right)+c_{44}\left(\partial u_{3}^{i} / \partial \xi_{2}\right)+2 c_{344} \varepsilon_{23}^{i}, \\
		&A_{35}=A_{53}=c_{33}\left(\partial u_{1}^{i} / \partial \xi_{3}\right)+c_{55}\left(\partial u_{3}^{i} / \partial \xi_{1}\right)+2 c_{355} \varepsilon_{31}^{i}, \\
		&A_{45}=A_{54}=c_{44}\left(\partial u_{1}^{i} / \partial \xi_{2}\right)+c_{55}\left(\partial u_{2}^{i} / \partial \xi_{1}\right)+2 c_{456} \varepsilon_{12}^{i}.
	\end{aligned}
\end{equation}


For an initially isotropic body subjected to a uniform triaxial strain field, the velocity $v_{p}$ of a longitudinal wave propagating in the $\xi_3$ direction is given by \cite{Pao1985AcoustoelasticWI}:  
\begin{equation}
\rho^{o} v_{p}^{2} = c_{33} + (c_{31} + c_{331}) \varepsilon_{11}^{i} + (c_{32} + c_{332}) \varepsilon_{22}^{i} + (3c_{33} + c_{333}) \varepsilon_{33}^{i}  .
\end{equation}

For isotropic media, the coefficients are given as $c_{31} = \lambda$, $c_{32} = \lambda$, $c_{33} = \lambda + 2\mu$, $c_{331} = 2l$, $c_{332} = 2l$, and $c_{333} = 2l + 4m$, where $\lambda$ and $\mu$ denote the second-order moduli, and $l$, $m$, and $n$ represent the Murnaghan third-order moduli. 

Under the condition of uniaxial stress $\sigma$ applied in the $\xi_3$ direction, $\varepsilon^i_{11} = \varepsilon^i_{22} = -v \sigma / E$, $\varepsilon^i_{33} = \sigma / E$, where the Poisson’s ratio $v = \lambda / [2(\lambda + \mu)]$, and the Young’s modulus $E = [(3\lambda + 2\mu) \mu] / (\lambda + \mu)$. Thus, the longitudinal wave propagating along the direction of stress in the natural frame of reference can be expressed as:  
\begin{equation} \label{eq:wave_longitudinal}
	\rho^{o} v_{p}^{2} = \lambda + 2\mu + \frac{\sigma}{3\lambda + 2\mu} \left[ \frac{\lambda + \mu}{\mu} (2\lambda + 6\mu + 4m) + \lambda + 2l \right].  
\end{equation} 

The wave equation of the longitudinal wave propagating along the stress direction in the natural frame of reference is 
\begin{equation} \label{eq:wave_equation}
	\left\{\begin{aligned}
		&p \frac{\partial^{2}{u}}{\partial t^{2}}-\nabla^2u={f}; \\
		&{u}(x,0)=0; \\
		&\frac{\partial{u}(x, 0)}{\partial t}=0; \\
		&\left.u(x,t)\right|_{\Gamma}= 0.
	\end{aligned}\right.
\end{equation}
where $x$  in a domain $\Omega$ with boundary $\Gamma$ , and $p = 1/v_{p}^{2}$.  

According to the research by Egle and Bray \cite{egle1976measurement}, the acoustoelastic effect of longitudinal waves propagating parallel to the stress direction is the most significant, showing far greater sensitivity to stress compared to other wave types. The primary residual stresses in welded specimens are caused by tensile/compressive effects. It is acceptable to determine welding residual stresses assuming a mainly uniaxial effect. In general, this uniaxial residual stress is distributed inhomogeneously over the cross-section of the component \cite{radaj2012heat, leon1996residual}. Therefore, it is reasonable to use Eq.~(\ref{eq:wave_longitudinal}) and Eq.~(\ref{eq:wave_equation}) for the inversion of non-uniform residual stresses in welded specimens.

\subsection{Inversion theory}
Based on the above acoustoelasticity theory, the following derivation of the nonuniform residual stress inversion theory is derived. Nonuniform residual stress inversion is reformulated as an optimization problem. In this study, the equation \( g(x,p) = 0 \) describes the relationship between displacement and residual stress, where \( p \) represents the nonuniform residual stress, and \( x \) corresponds to the displacement at various points. The goal is to infer nonuniform residual stress from measured displacements, a classic inverse problem. The solution to this problem involves the following steps \cite{CUICongyue_246}:  (1) Define an objective function \( J(p) \) to measure the discrepancy between numerical simulation predictions and observed data; (2) Compute the gradient of the objective function, \( \partial J / \partial p \); (3) Apply optimization techniques, such as gradient descent, to minimize \( J(p) \) and estimate the parameters \( p \).

\subsubsection{The objective function}
The objective function is used to describe the misfit between the predicted records of numerical simulation and the actual observed records. The least-squares functional is chosen as the objective function:  
\begin{equation}\label{eq:loss_func}
	J(p)=\frac{1}{2}\sum_{i=1}^{n_{s}} \int_0^T  (u_{i}^{pred}(p)-u_{i}^{obs})^{2}  dt,
\end{equation}
Eq.(\ref{eq:loss_func}) can also be written as follows,
\begin{equation}
	J(p)=\int_0^T\int_\Omega\frac{1}{2}\sum_{i=1}^{n_{s}} (u_{i}^{pred}(p)-u_{i}^{obs})^{2} \delta (x - x_r) d\Omega dt.
\end{equation}
where $T$ is the recording time, $n_{s}$ is the number of receivers. $u^{pred}$ is the predicted numerical simulation record, which depends on the parameter $m$ we choose. $u^{obs}$ is the actual observed record,  $x_r$ are the coordinates of the position of the receivers. 
\subsubsection{Adjoint method}
The adjoint method is a technique for the efficient calculation of the gradient of the functional which is to be minimized in
the solution process \cite{GIVOLI2021113810,Fichtner2011FullSW}. The adjoint method is used to compute the gradient of the objective function $J(p)$ with respect to the parameter $p$. Use the Lagrangian to formulate the adjoint method:  
\begin{equation}
	\begin{aligned}
		\mathcal{L}&=\int_0^T\int_\Omega\frac{1}{2}\sum_{i=1}^{n_{s}} (u_{i}^{pred}(p)-u_{i}^{obs})^{2} \delta (x - x_r) d\Omega dt \\
		&+ \int_0^T\int_\Omega w (p\frac{\partial^{2}{u}}{\partial t^{2}}-\nabla^2u - {f}) d\Omega dt.
	\end{aligned}
\end{equation}
where $w$ is a function of time. Since the wave equation \(p\frac{\partial^{2}{u}}{\partial t^{2}}-\nabla^2u - {f} = 0\) is satisfied at any moment \(t\) on the domain \(\Omega\), we get
\begin{equation}
	\frac{\partial J}{\partial p} = \frac{\partial \mathcal{L}}{\partial p},
\end{equation}

The variation of \(\mathcal{L}\) is obtained by taking the variations of the objective function term and the wave equation term with respect to \(u\) and \(p\), and then summing them.
\begin{equation}
	\begin{aligned}
		\delta \mathcal{L} &= \int_0^T\int_\Omega \sum_{i=1}^{n_{s}} (u_{i}^{pred}(p)-u_{i}^{obs})\delta (x - x_r) \delta u d\Omega dt  \\
		&+ \int_0^T\int_\Omega w\left(\frac{\partial^{2}{u}}{\partial t^{2}} \delta p + p \frac{\partial^{2}{\delta u}}{\partial t^{2}} -  \nabla^2 \delta u\right)d\Omega dt,
	\end{aligned}
\end{equation}
For the $w \cdot p \partial_{t}^{2} \delta{u}$, integrate by parts twice, 
\begin{equation}
	\begin{aligned}
		&\int_0^T\int_\Omega w \cdot p\frac{\partial^2 \delta u}{\partial t^2}d\Omega dt \\
		&=\int_0^T\int_\Omega \delta u \cdot p \frac{\partial^2w}{\partial t^2}d\Omega dt 
		+\int_\Omega\left(\left.w \cdot p \frac{\partial \delta u}{\partial t}\right|_{t=0}^{t=T}-\left.\delta u \cdot p \frac{\partial w}{\partial t}\right|_{t=0}^{t=T}\right)d\Omega, \\
		&=\int_0^T\int_\Omega \delta u \cdot p \frac{\partial^2w}{\partial t^2}d\Omega dt 
		+\int_\Omega\left(\left.w \cdot p \frac{\partial \delta u}{\partial t}\right|_{t=T}-\left.\delta u \cdot p \frac{\partial w}{\partial t}\right|_{t=T}\right)d\Omega. \\
	\end{aligned}
\end{equation}
For the $w \nabla^2 \delta u$, applying the Divergence Theorem \cite{10.1785/0120060041}, 
\begin{equation}\label{eq:Divergence1}
	\int_\Omega w\nabla^2 \delta u d\Omega 
	= \int_{\Gamma }\hat{n}\cdot\left\{ w\nabla \delta u \right\}d\Gamma - \int_\Omega \nabla w : \nabla \delta u d\Omega  .\\
\end{equation}
where “:” denotes tensor double contraction, and similarly
\begin{equation}\label{eq:Divergence2}
	\int_\Omega \delta u \nabla^2 w d\Omega 
	= \int_{\Gamma }\hat{n}\cdot\left\{ \delta u\nabla w \right\}d\Gamma - \int_\Omega \nabla \delta  u : \nabla w d\Omega  .\\
\end{equation}
Subtracting Eq.~(\ref{eq:Divergence2}) from Eq.~(\ref{eq:Divergence1}) gives:
\begin{equation}
	\int_\Omega w\nabla^2 \delta u d\Omega = \int_\Omega \delta u\nabla^2 w d\Omega +  \int_{\Gamma}\left(w\frac{\partial \delta u}{\partial n}- \delta u\frac{\partial w}{\partial n}\right)d\Gamma .
\end{equation}

The variation of the objective functional \(J\) is given by the sum of five integral terms:
	\begin{equation}\label{eq:grad_equation}
		\begin{aligned}
			\delta J&=\int_0^T\int_\Omega \left(w\frac{\partial^{2}{u}}{\partial t^{2}}\right) \delta p d\Omega dt  \\
			& +  \int_0^T\int_\Omega \left[p \frac{\partial^2w}{\partial t^2} -  \nabla^2w + \sum_{i=1}^{n_{s}} (u_{i}^{pred}-u_{i}^{obs})\delta (x - x_r)\right] \delta u d\Omega dt  \\
			& + \int_\Omega \left.w \cdot p \frac{\partial \delta u}{\partial t}\right|_{t=T} d\Omega \\
			& - \int_\Omega  \left.\delta u \cdot p \frac{\partial w}{\partial t}\right|_{t=T} d\Omega \\
			& - \int_0^T \int_{\Gamma}w\frac{\partial \delta u}{\partial n} d\Gamma dt .\\
		\end{aligned}
	\end{equation}
	Here, the first term reflects the coupling between the test function \(w\) and the second time derivative of the state variable \(u\) through the variation \(\delta p\). 

To obtain \(\delta J / \delta p\), the last four terms of Eq.~(\ref{eq:grad_equation}) are set to zero, the adjoint equation is derived:
	\begin{align}
		p \frac{\partial^2 w}{\partial t^2} - \nabla^2 w 
		+ \sum_{i=1}^{n_s} \left(u_i^{\text{pred}} - u_i^{\text{obs}}\right) \delta(x - x_r) = 0.
\end{align}

The derivation also yields natural boundary and end conditions.
	End conditions at \(t = T\) ensure fixed states at the end time.
	\begin{align}
		w(T) = 0, \quad \frac{\partial w}{\partial t}(T) = 0,
	\end{align}
	Boundary conditions on the spatial domain boundary \(\Gamma\) reflect zero-value constraints on the boundary.
	\begin{align}
		w = 0 \quad \text{on} \quad \Gamma.
\end{align}

So the complete adjoint equation is obtained:
\begin{equation} \label{eq:adjoint_equation}
	\left\{\begin{aligned}
		&p \frac{\partial^2w}{\partial t^2} -  \nabla^2w = \sum_{i=1}^{n_{s}} (u_{i}^{obs} - u_{i}^{pred})\delta (x - x_r); \\
		&w(T)= 0; \\
		&\frac{\partial w(T)}{\partial t} = 0;\\
		&\left.w(x,t)\right|_{\Gamma}= 0.
	\end{aligned}\right.
\end{equation}

The wave equation (Eq.~(\ref{eq:wave_equation})) describes the forward propagation of wavefields, while the adjoint equation (Eq.~(\ref{eq:adjoint_equation})) is solved backward in time to compute the gradient of \( J \). These two equations share a similar form, with the key difference being the source term. Specifically, \( \partial J / \partial p \) is obtained by solving the wave equation one more time, but in reverse time.

To further clarify the adjoint equation (Eq.~(\ref{eq:adjoint_equation})), we introduce a time-reversal transformation:
\begin{equation}
	q(t) = w(T - t),
\end{equation}
which helps simplify the adjoint equation and makes the time-reversal structure more evident. Under this transformation, the adjoint equation becomes:
\begin{equation} \label{eq:adjoint_equation_2}
	\left\{
	\begin{aligned}
		&p \frac{\partial^2 q}{\partial t^2} -  \nabla^2 q = \sum_{i=1}^{n_{s}} \left(u_{i}^{obs}(T - t) - u_{i}^{pred}(T - t)\right)\delta(x - x_r); \\
		&q(0) = 0; \\
		&\frac{\partial q(0)}{\partial t} = 0; \\
		&\left.q(x,t)\right|_{\Gamma} = 0.
	\end{aligned}
	\right.
\end{equation}

The primary difference between Eq.~(\ref{eq:wave_equation}) and Eq.~(\ref{eq:adjoint_equation_2}) lies in the source term. The term \( u^{obs}(T - t) - u^{pred}(T - t) \) represents the discrepancy between observed and predicted waveforms, evaluated in reverse time, as discussed by \cite{j.1365-246X.2006.02978.x}. This transformation allows us to solve for the adjoint state \( q(x,t) \), which is then used to compute the gradient of the loss functional.

Once \( q(x,t) \) is obtained, the gradient of the objective functional \( J \) with respect to the parameter \( p \) is computed using the following expression:
\begin{equation}
	\frac{\partial J}{\partial p}(x) = \sum_{i=1}^{n_{s}} \int_0^T q(x,T-t)\frac{\partial^2 u(x,t)}{\partial t^2} dt.
\end{equation}

\subsection{Inverse flowchart}
\begin{figure}[ht]
	\centering 
	\includegraphics[width=0.5\columnwidth]{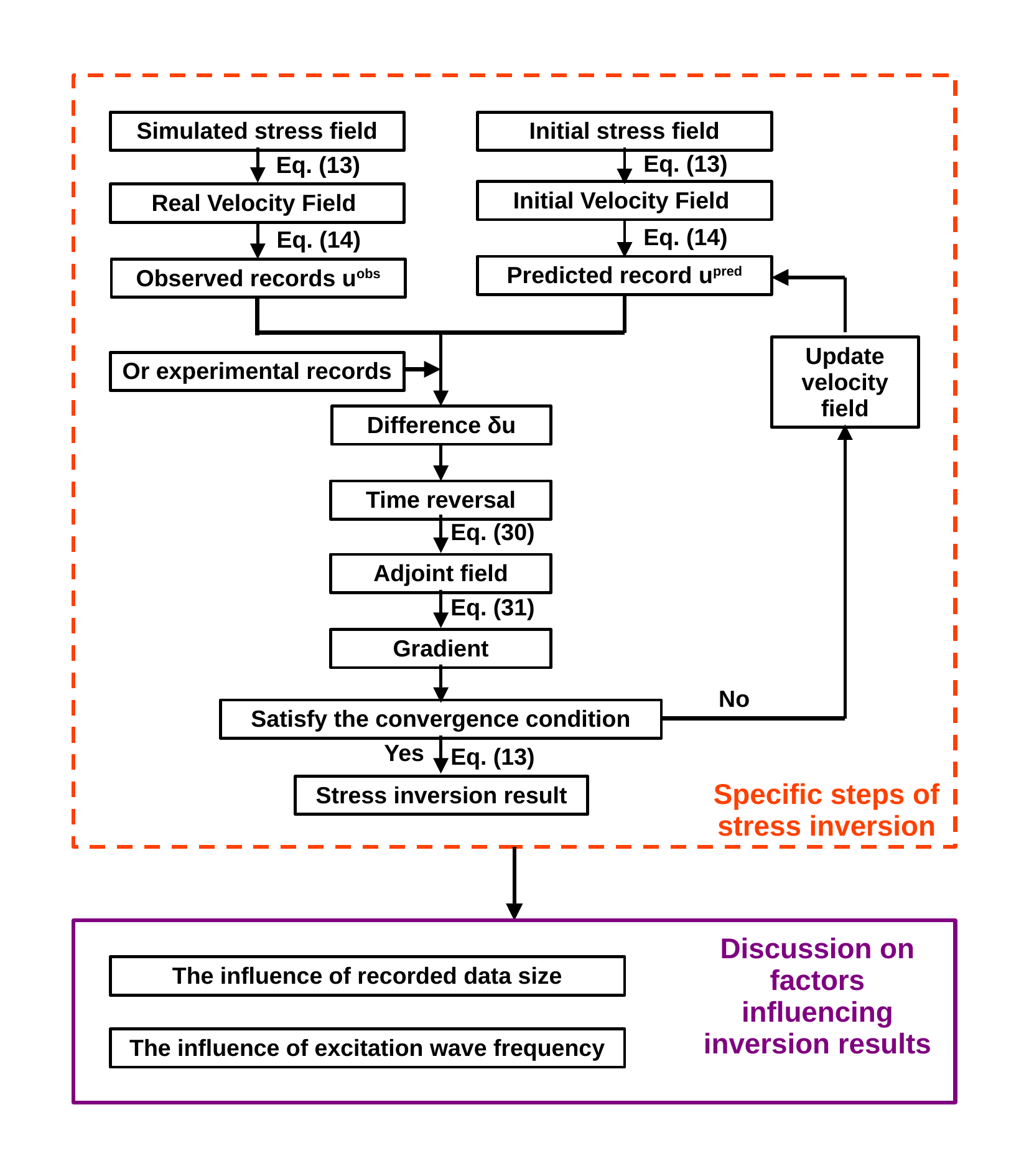} 
	\caption{Flowchart of residual stress inversion} 
	\label{fig:inverse_map}
\end{figure}

To provide a clearer understanding of the inversion process, Fig.~\ref{fig:inverse_map} illustrates the inversion flowchart. The flowchart is organized into two primary components. The first part outlines the specific steps of stress inversion, with equations provided to guide each step of the process. A fixed step size is employed for stress updates. The second part examines factors influencing inversion results, focusing on two key aspects: the size of recorded data and the frequency of excitation waves.

\section{Numerical simulation}
\subsection{Welding simulation}
\begin{figure}[ht]
	\centering 
	\includegraphics[width=0.5\columnwidth]{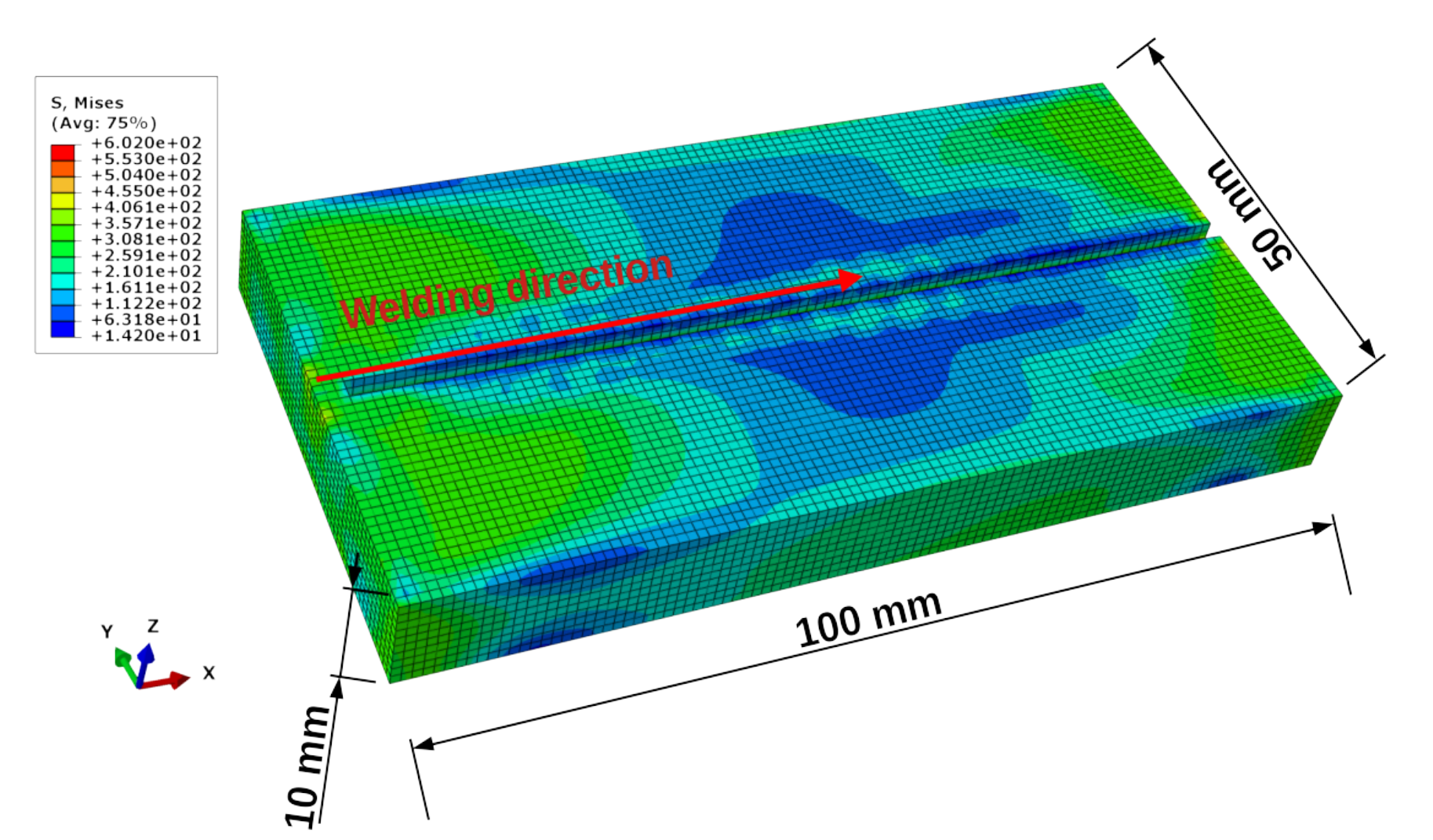} 
	\caption{Three-dimensional dimensions of welded specimen} 
	\label{fig:3D}
\end{figure}

\begin{figure}[ht]
	\centering 
	\includegraphics[width=0.5\columnwidth]{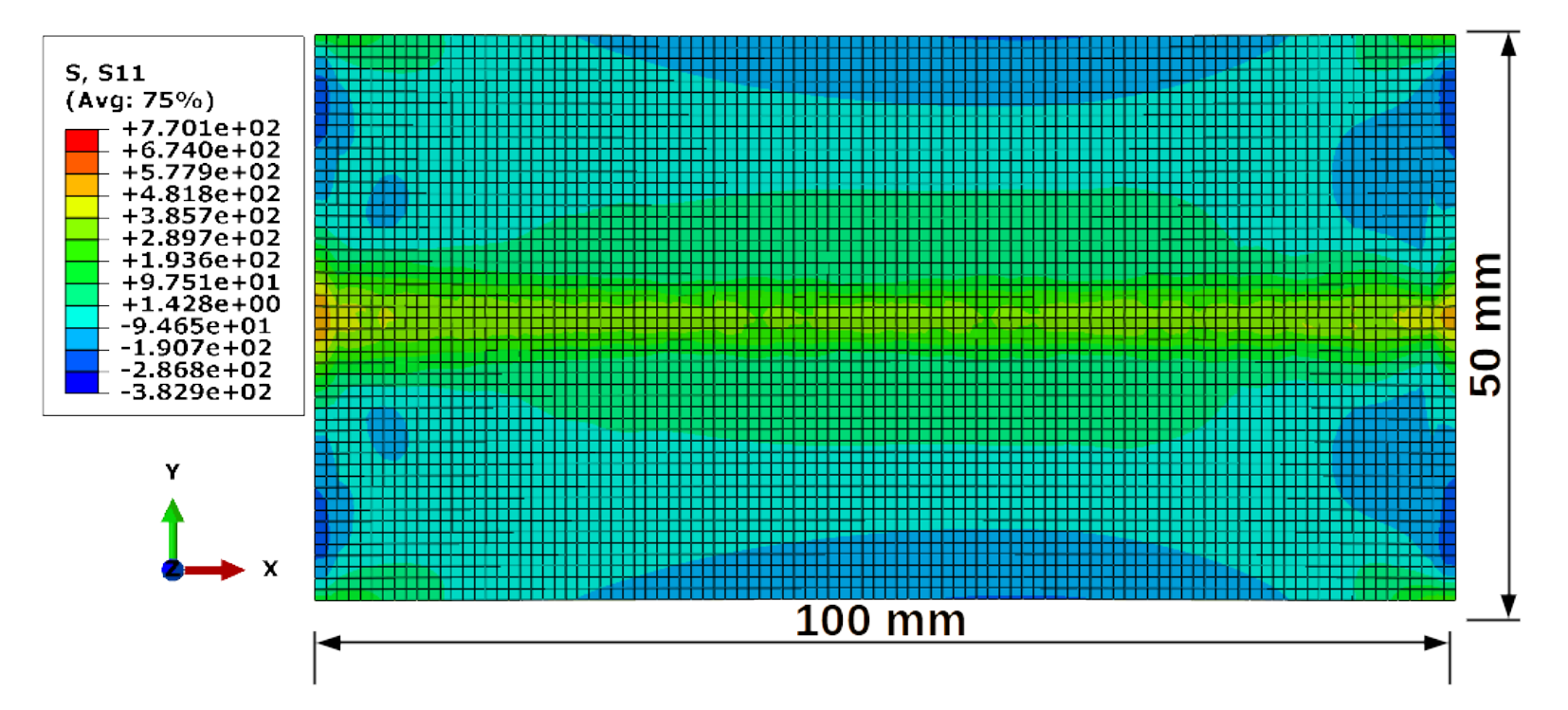} 
	\caption{Distribution of longitudinal residual stress in welded specimen (MPa)} 
	\label{fig:S11}
\end{figure}

\begin{figure}[ht]
	\centering 
	\includegraphics[width=0.5\columnwidth]{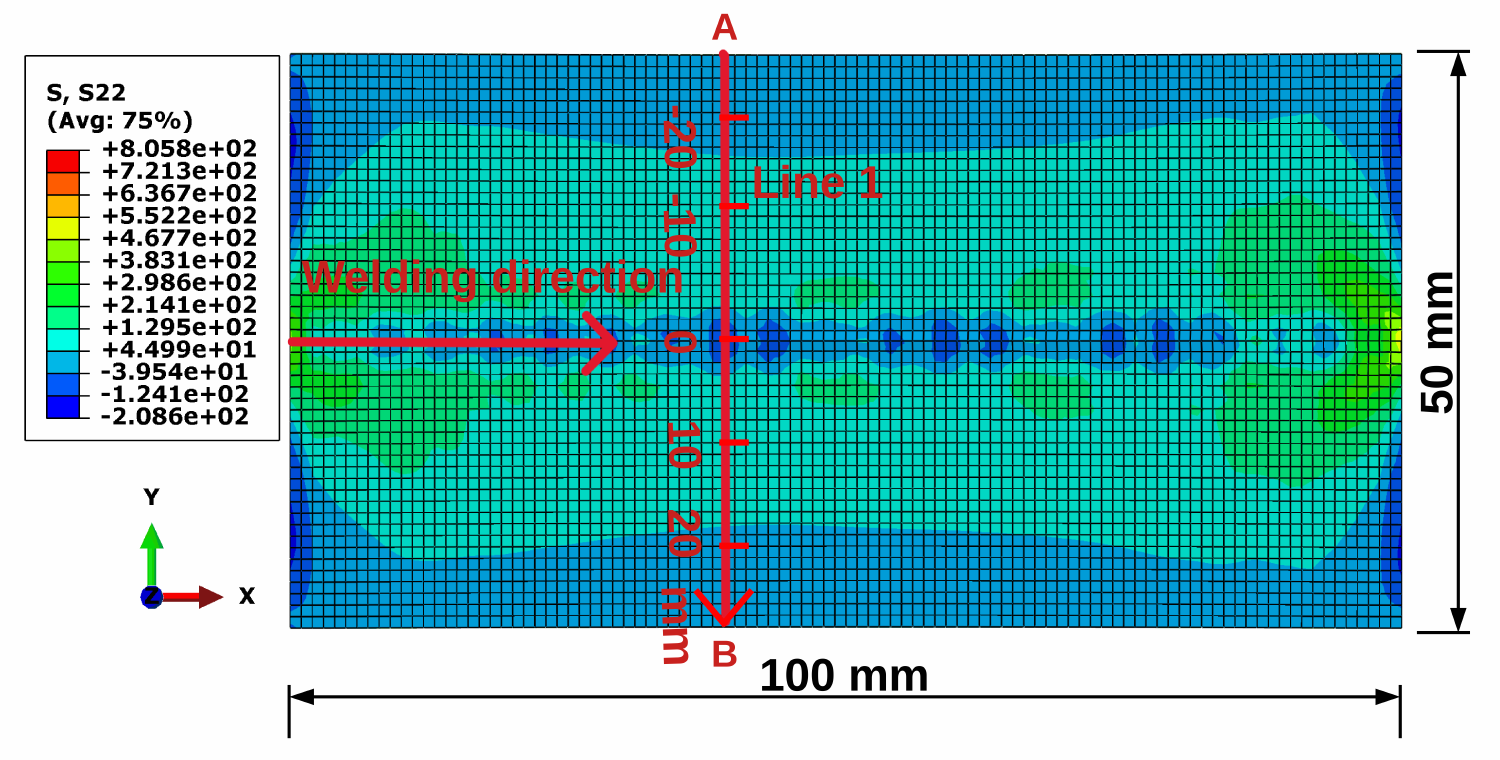} 
	\caption{Distribution of transverse residual stress in welded specimen (MPa)} 
	\label{fig:S22}
\end{figure}

Measuring the residual stress of welding material is a typical application of the acoustoelastic theory \cite{WANG2018117,LUO2023102765}. In order to better demonstrate the inversion effect, we use finite element method to simulate the welding process \cite{ILKERYELBAY201029,JAVADI2013628}. The three-dimensional dimensions of the material are 100 $\times$ 50 $\times$ 10 mm. The weld is applied on the upper surface of the specimen, with the weld center positioned at the midpoint of the specimen's width. The welding direction is along the length of the specimen, with a weld length of 100 mm and a weld width of 2 mm. The schematic diagram of the welded specimen is shown in Fig.~\ref{fig:3D}. There are no boundary constraints in this model. The welding heat source is double ellipsoid source. Transverse residual stress is the stress in the direction perpendicular to the welding direction, while longitudinal residual stress is the stress along the welding direction, both formed due to thermal expansion and contraction constraints during welding. The welding material is 304 stainless steel. The longitudinal and transverse residual stress distributions on the upper surface of the specimen are shown in Fig.~\ref{fig:S11} and Fig.~\ref{fig:S22}. 


\begin{figure}[ht]
	\centering 
	\includegraphics[width=0.5\columnwidth]{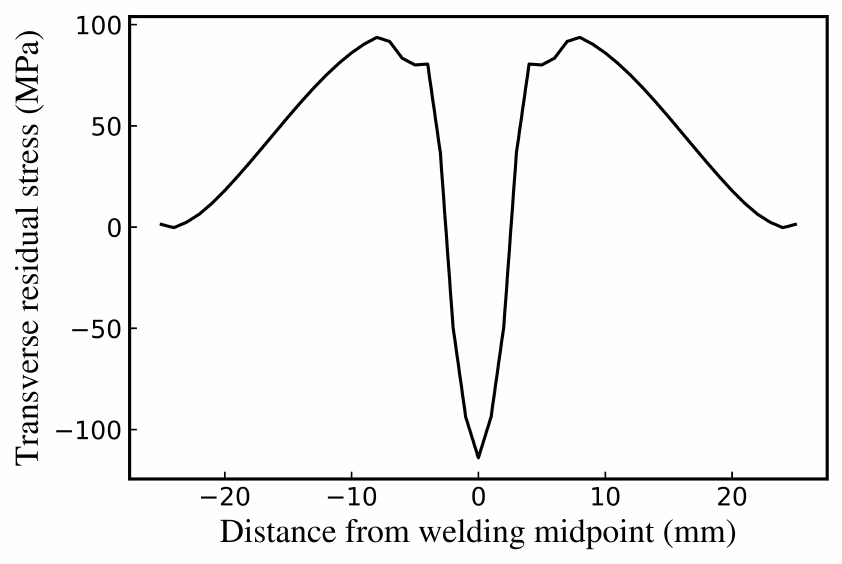} 
	\caption{The transverse residual stress distribution on line 1} 
	\label{fig:line1_S22}
\end{figure}

Select a line segment (line 1) perpendicular to the welding direction on the upper surface of the specimen, with endpoints A and B. Taking the welding center as the origin, a coordinate system is established along the direction perpendicular to the welding direction, as shown in Fig.~\ref{fig:S22}. The coordinates of points A and B are \(-25 \, \text{mm}\) and \(+25 \, \text{mm}\), respectively. Along the established coordinate system, the transverse residual stress distribution on line 1 is shown in Fig.~\ref{fig:line1_S22}.

In this paper, inhomogeneous anisotropy caused by crystallographic and morphological texture in the weldment and the surrounding heat-affected zone is not considered \cite{lu2008ultrasonic, joseph2015evaluation}. 

\subsection{Ultrasonic simulation}
\begin{figure}
	\centering 
	\includegraphics[width=0.5\columnwidth]{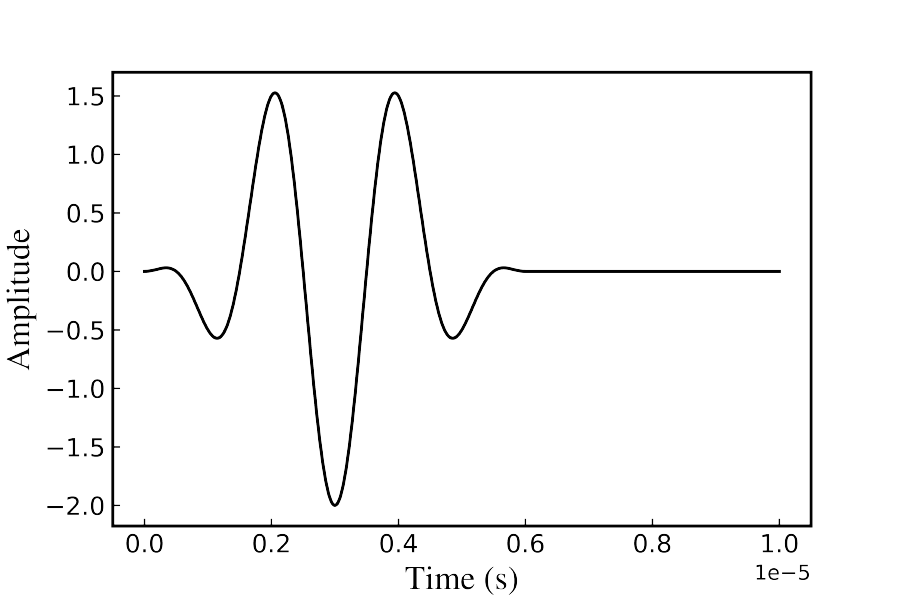} 
	\caption{Time domain curve of ultrasonic excitation wave} 
	\label{fig:incident_wave}
\end{figure}

The finite difference method is used to simulate the propagation process of longitudinal waves in stress. Simulation parameters: time step size is 1ns, the total time is 10000 ns, the space step size is 0.1 mm, the propagation distance is 5 cm. The material properties of steel are shown in Table \ref{table:Fe_TOEC} \cite{Blaschke2017}. The incident ultrasonic wave is modeled by the transient excitation pulse. The equivalent force signal is given in the Eq.~(\ref{eq:ultrasonic_wave}) and 0.5 MHz is selected as the frequency parameter \cite{baskaran2007simulation}. The time-domain diagram of the excitation signal is shown in Fig.~\ref{fig:incident_wave}.
\begin{table}[ht]
	\centering
	\caption{The density, Lamé constants and Murnaghan constants of steel}
	\resizebox{0.8\columnwidth}{!}{\begin{tabular}{cccccc}
			\hline
			$\rho^o$ & $ \lambda$ & $ \mu$ & ${\rm l}$ & ${\rm m}$ & ${\rm n}$ \\ \hline
			7860 ${\rm kg/m^3}$ & 110.8 ${\rm GPa}$ & 86.8 ${\rm GPa}$ & –345 ${\rm GPa}$ & –632 ${\rm GPa}$ & –660 ${\rm GPa}$  \\ \hline
	\end{tabular}}
	\label{table:Fe_TOEC}
\end{table}
\begin{equation}\label{eq:ultrasonic_wave}
	F(t) = 
	\begin{cases}
		\left[1-\cos(\frac{2\pi f}{3}t)\right] \cos(2\pi ft), \quad &\text{for} \quad 0 \le t \le \frac{3.0}{f}\\
		0. &\text{otherwise}
	\end{cases}
\end{equation}

\section{Results and discussion}
\subsection{Influence of the number of sources and receivers}
\begin{figure*}[htbp]
	\centering
	\subfloat[one receiver: point B, position of receiver: 25 mm]{\label{subfig:1_source_a}\includegraphics[width = 0.5\columnwidth]{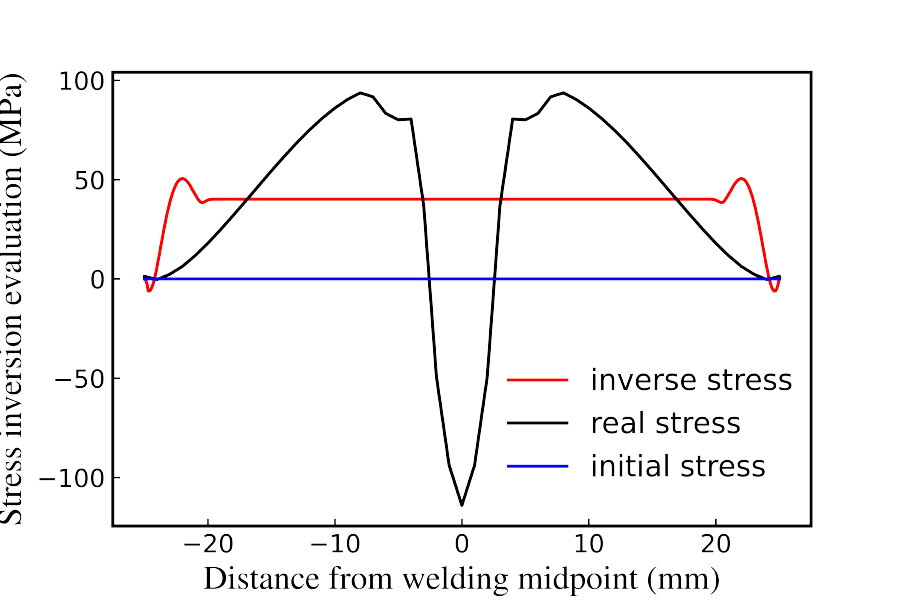}}
	\subfloat[two receivers, position of receivers: 0, 25 mm]{\includegraphics[width = 0.5\columnwidth]{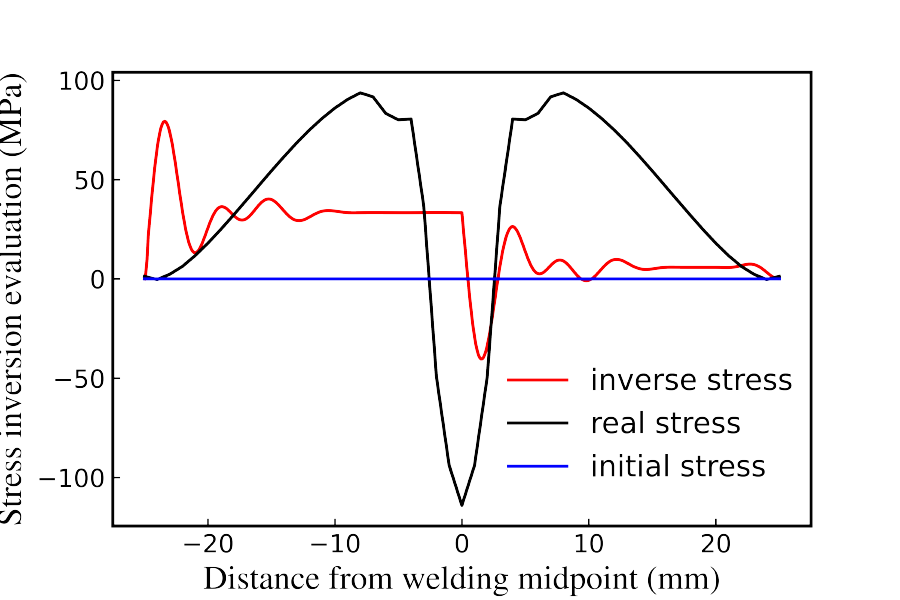}} \\
	\subfloat[three receivers, position of receivers: -8.4, 8.2, 25 mm]{\includegraphics[width = 0.5\columnwidth]{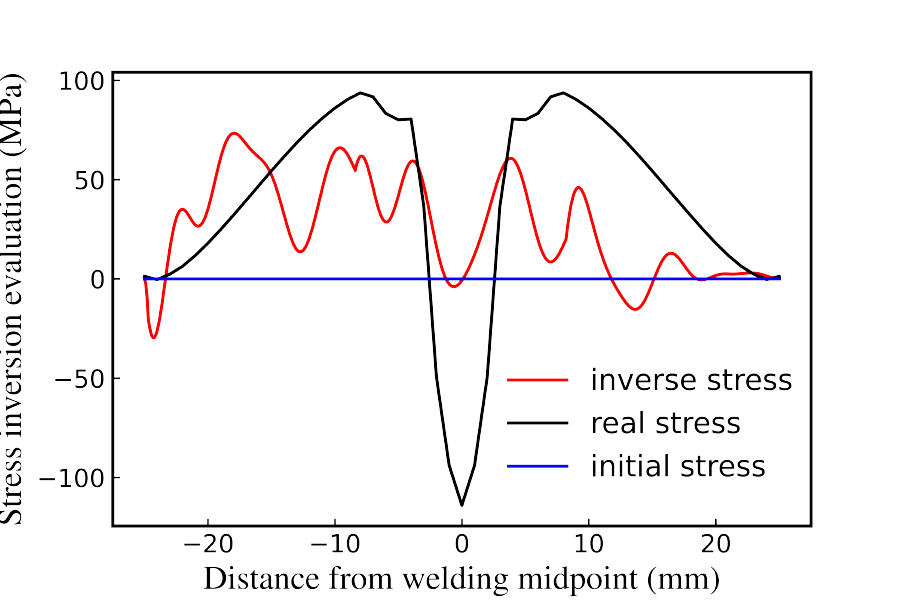}}
	\subfloat[four receivers, position of receivers: -12.5, 0, 12.5, 25 mm]{\includegraphics[width = 0.5\columnwidth]{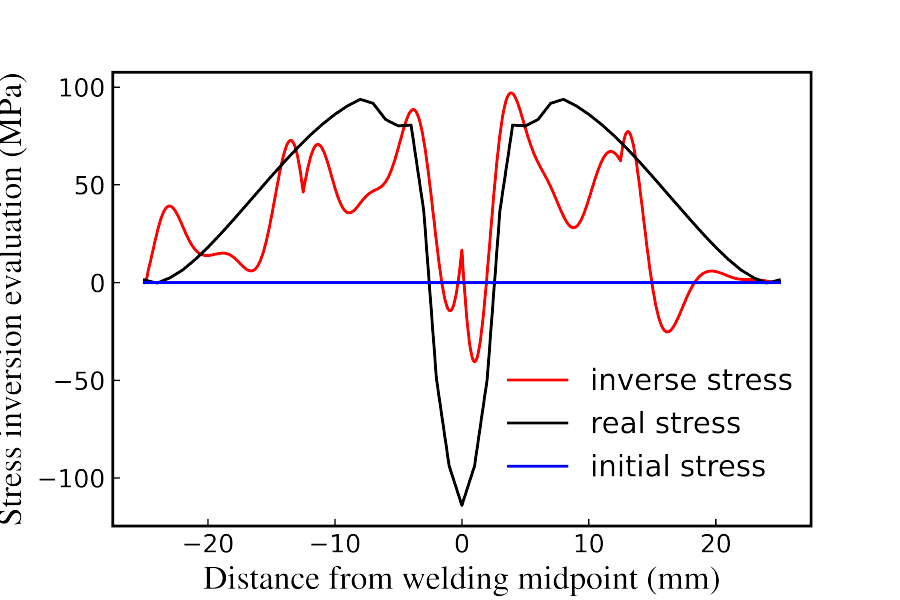}} \\
	\subfloat[five receivers, position of receivers: -15, -5, 5, 15, 25 mm]{\includegraphics[width =0.5\columnwidth]{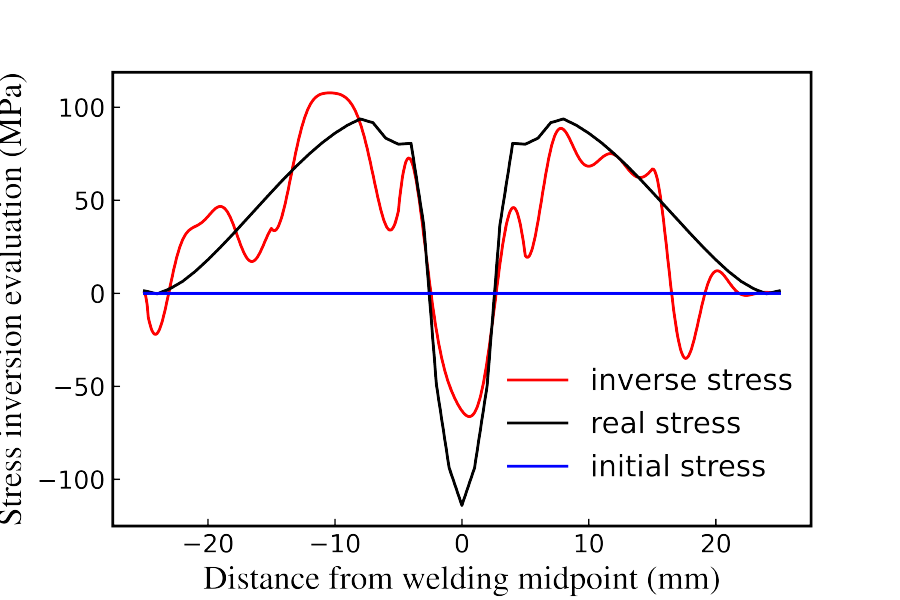}}
	\subfloat[six receivers, position of receivers: -15, -5, 0, 5, 15, 25 mm]{\includegraphics[width = 0.5\columnwidth]{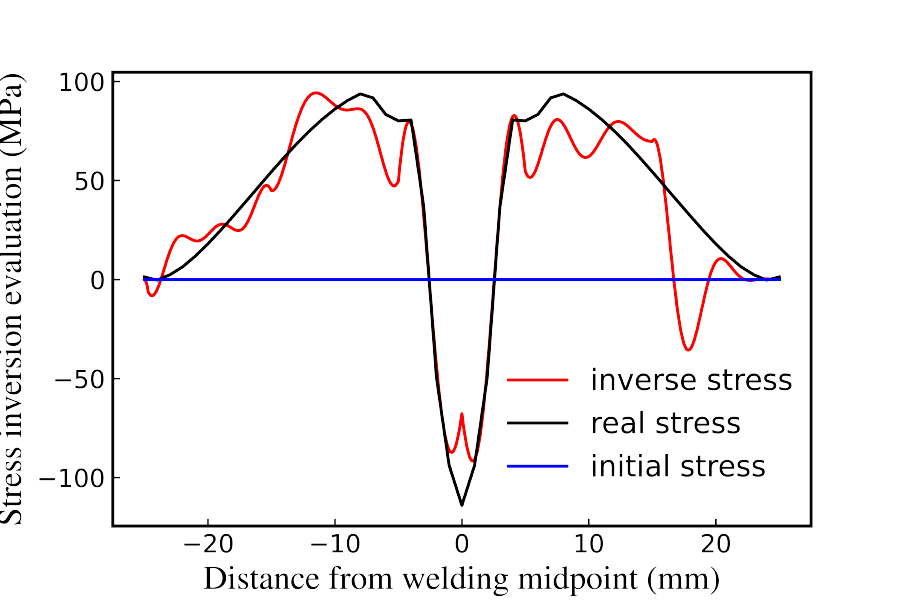}} 
	\caption{Inversion result under one source and different number of receivers}
	\label{fig:1_source}
\end{figure*}

The ultrasonic wave is excited at point A and received at point B, the ultrasound propagates from point A to point B along line 1. The excitation and reception positions are defined according to the coordinate system in Fig.~\ref{fig:S22}. The initial stress is selected as 0 MPa, and the residual stress inversion is performed according to the inversion map shown in Fig.~\ref{fig:inverse_map}. The result of stress inversion based on data obtained from the excitation point A and the receiving point B is shown in Fig.~\ref{subfig:1_source_a}, but the inversion result is meaningless obviously. This highlights a common characteristic of inverse problems: limited information often leads to incorrect inversion results. To address this, the ultrasonic wave is excited at point A, while the number of receiving points is increased to gather more information. Inversion results for varying numbers of receivers are shown in Fig.~\ref{fig:1_source}. From the stress inversion results, in the case of one excitation point, the inversion results of three or less receivers are wrong and are irrelevant to the real stress distribution. With four or more receiving points, the residual stress distribution trend becomes reliable. When the number of receiving points is six, the results of residual stress inversion near the welding center are very accurate, but the results far away from the welding center are poor.

\begin{figure*}[htbp]
	\centering
	\subfloat[six sources, position of sources: -25, -15, -5, 0, 5, 15 mm; \protect\\ one receivers, position of receivers: 25 mm]{
		\includegraphics[scale=0.6]{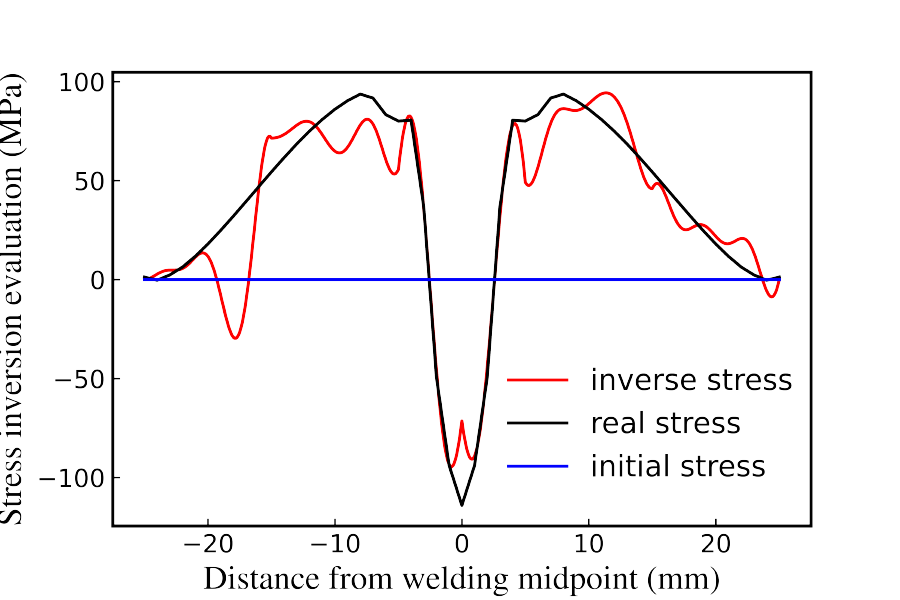}
	}
	\subfloat[six sources, position of sources: -25, -15, -5, 0, 5, 15 mm; \protect\\ six receivers, position of receivers: -15, -5, 0, 5, 15, 25 mm]{
		\includegraphics[scale=0.6]{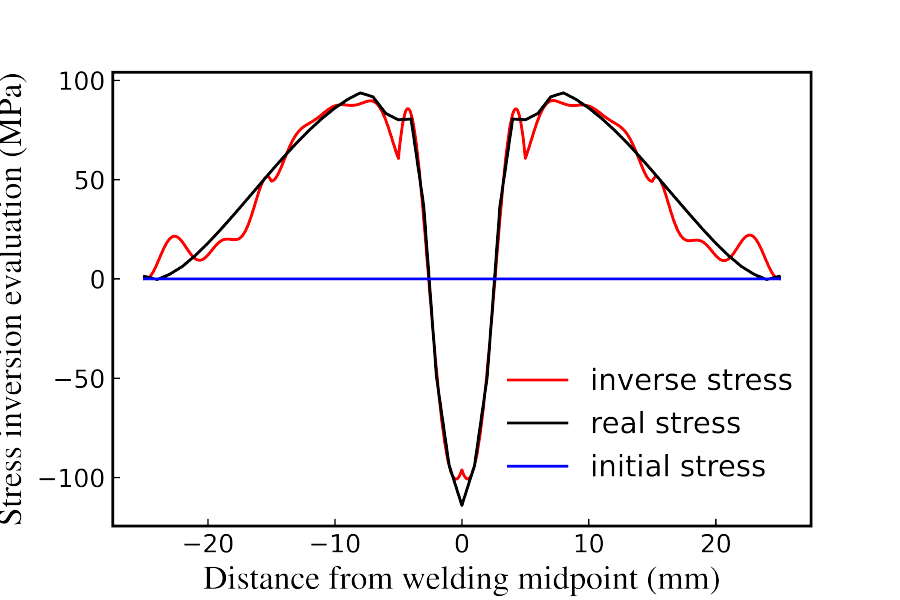}
	}
	\caption{Inversion result under six sources and different number of receivers}
	\label{fig:6_and_6_source}
\end{figure*}

To improve the accuracy of inversion results, the number of sources is increased, and the corresponding residual stress inversion results are presented in Fig.~\ref{fig:6_and_6_source}. As the number of sources increases, the residual stress inversion results show greater agreement with the actual stress distribution, particularly in regions farther from the welding center.

\subsection{Influence of excitation wave frequency}
\begin{figure}[ht]
	\centering
	\subfloat[Excitation wave frequency: 1MHz]{\includegraphics[scale=0.6]{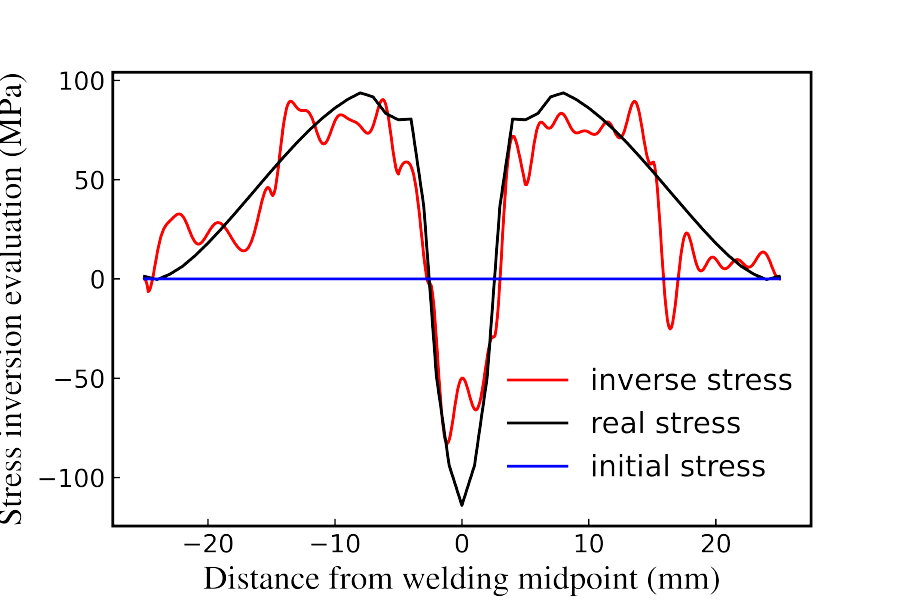}}
	\subfloat[Excitation wave frequency: 1.5MHz]{\includegraphics[scale=0.6]{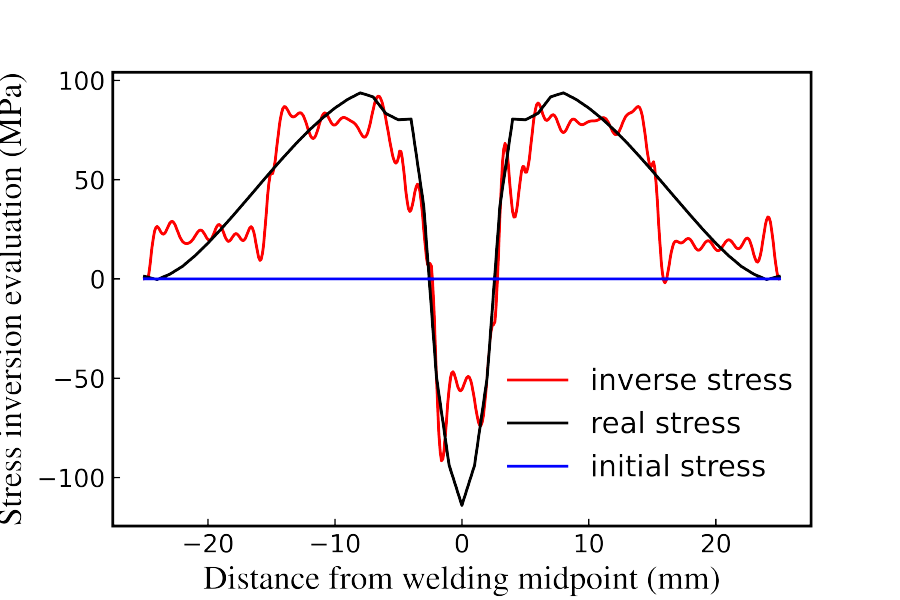}} \\
	\subfloat[Excitation wave frequency: 2.5MHz]{\includegraphics[scale=0.6]{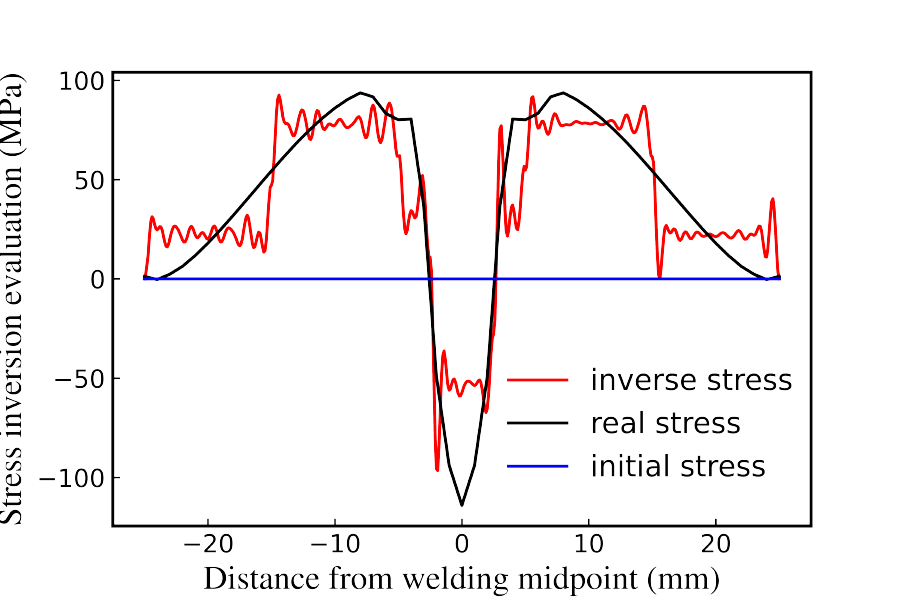}}
	\subfloat[Excitation wave frequency: 5MHz]{\includegraphics[scale=0.6]{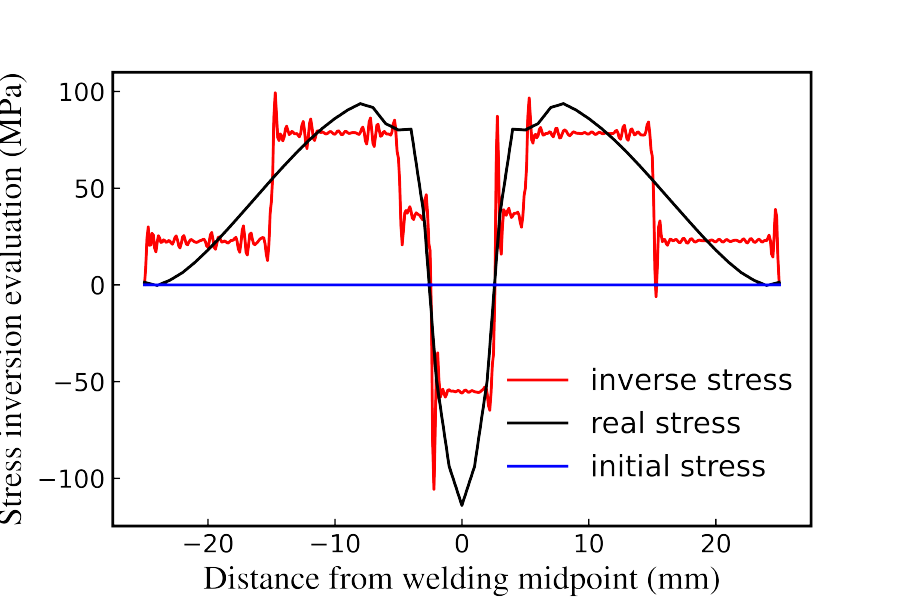}} 
	\caption{Inversion result under different excitation wave frequencies (one source, position of source: -25 mm; seven receivers, position of receivers: -15, -5, -2.5, 2.5, 5, 15, 25 mm)}
	\label{fig:1_1.5_2.5_5MHz}
\end{figure}

According to studies in FWI \cite{10.1190/1.3238367}, inversion is challenging when the frequency of the excitation wave is high. Residual stress inversion is performed by varying the excitation wave frequency. The inversion results deteriorate as the frequency increases, and the results indicate that high-frequency excitation waves still pose challenges for accurate residual stress inversion.  The inversion results for excitation wave frequencies of 1 MHz, 1.5 MHz,  2.5 MHz and 5 MHz are shown in Fig.~\ref{fig:1_1.5_2.5_5MHz}. As the excitation wave frequency increases, the residual stress inversion results worsen, particularly in the welding center, which typically corresponds to the region of highest stress and the area of greatest interest. At the excitation wave frequency of 5 MHz, the inversion results exhibit nearly horizontal line segments between neighboring receiving points, indicating a significant discrepancy from the true stress distribution.

High frequency waves are typically suitable for detecting small defects and fine stress variations near the surface, providing high spatial resolution. However, The inversion is inherently an optimization problem, and during this process, it may converge to local minima. In regions with stress concentration, where stress varies sharply, inversion algorithms are more prone to being trapped in local minima, resulting in inaccurate physical parameter estimates. To achieve better inversion results, the multiscale inversion strategy can be adopted—starting with low-frequency data to stably estimate the overall structure, and then progressively incorporating high-frequency data to refine the stress and physical parameters in localized regions\cite{10.1190/1.3238367}.

\subsection{Experimental roadmap}
To validate the effectiveness of the residual stress detection inversion method proposed in this paper, the subsequent experimental procedure is designed. The specific experimental steps are as follows:
\begin{enumerate}
\item Sample Preparation: Two metal samples of identical material and dimensions are selected. One will be subjected to welding to induce residual stresses, while the other will be used to determine the acoustoelastic coefficient.
\item Pre-Welding Testing: Prior to welding, the first sample is selected, and reference points are marked along the transverse cross-section perpendicular to the weld. The length of this cross-section is divided into six equal parts, and seven equidistant reference points are marked (one excitation point and six receiving points). Longitudinal ultrasonic waves are excited at the excitation point, and displacement signals are recorded at the receiving points. This procedure is repeated for different excitation frequencies to obtain displacement curves at various excitation frequencies. These data will serve as the initial dataset.
\item Post-Welding Testing: After welding, the same testing procedure in step 2 will be performed at the previously marked points to collect experimental data for the welded sample.
\item Acoustoelastic Coefficient Determination: The tensile loading test is conducted on another specimen to determine its acoustoelastic coefficient.
\item Inversion Process: The inversion process follows the workflow outlined in Fig.~\ref{fig:inverse_map}. The misfit between the experimental data and initial data is computed, followed by the application of time reverse and gradient iteration to obtain the wave velocity distribution. Finally, the distribution of residual stress is calculated based on the measured acoustoelastic coefficient. 
\item Validation and Comparison: The residual stresses at the six receiving points will be measured using X-ray diffraction or hole-drilling methods. The inversion results will then be compared with the X-ray measurement results. The discussion will focus on the following aspects: (1) Qualitative analysis: Whether the inversion results align with the true stress distribution pattern; (2) Quantitative analysis: The discrepancy between the inversion stress values and the real stress values, along with an analysis of the sources of error; (3) Influencing factors: The impact of data volume and excitation frequency on the inversion accuracy.
\end{enumerate}

\section{Conclusion}
This study proposes a novel approach for residual stress evaluation, integrating acoustoelastic theory and full waveform inversion (FWI) concepts. Theoretical derivation and numerical simulations demonstrate the feasibility and reliability of this method. The results highlight several key findings and considerations for practical application.

First, the accuracy of residual stress inversion significantly depends on the amount of collected data. When the data is insufficient, the inversion results deviate considerably from the actual stress distribution. However, with sufficient data (specifically, at least six channels of information) the inversion results closely approximate the real residual stress distribution, proving the reliability of the proposed method.  

And then, the frequency of the excitation wave is a critical factor affecting inversion accuracy. High-frequency excitation waves lead to deterioration in the inversion results, particularly in regions of high stress concentration, such as the welding center. This limitation underscores the importance of selecting appropriate excitation wave frequencies for achieving precise stress characterization.  

In summary, the proposed residual stress evaluation approach offers a promising solution for high-precision stress distribution analysis, particularly in non-destructive testing scenarios. Future work will focus on experimental validation and optimization of the method under more complex conditions.

\section*{CRediT authorship contribution statement}
\textbf{Maoyu Xu}: Conceptualization, Methodology, Software, Validation, Formal analysis, Investigation, Writing -- original draft.  \textbf{Hongjian Zhao}: Data curation, Investigation. \textbf{Changsheng Liu}: Resources, Supervision, Funding acquisition. \textbf{Yu Zhan}: Visualization, Resources, Writing -- review \& editing, Funding acquisition.
\section*{Declaration of competing interest}
The authors declare that they have no known competing financial interests or personal relationships that could have appeared to influence the work reported in this paper.
\section*{Acknowledgements}
This study is supported by the National Natural Science Foundation of China Project (Grant No. 51771051), the Natural Science Foundation of Liaoning Province Project (Grant No. 2021-MS-102) and the Fundamental Research Funds for the Central Universities (Grant No. N2105021).
\section*{Data availability}
Data will be made available on request.
\bibliographystyle{ieeetr}
\bibliography{refs}

\begin{thebibliography}{10}

\bibitem{PhysRev.92.1145}
D.~S. Hughes and J.~L. Kelly, ``Second-order elastic deformation of solids,'' {\em Phys. Rev.}, vol.~92, pp.~1145--1149, Dec 1953.

\bibitem{thurston1964third}
R.~Thurston and K.~Brugger, ``Third-order elastic constants and the velocity of small amplitude elastic waves in homogeneously stressed media,'' {\em Physical Review}, vol.~133, no.~6A, p.~A1604, 1964.

\bibitem{Pao1985AcoustoelasticWI}
Y.~H. Pao and U.~Gamer, ``Acoustoelastic waves in orthotropic media,'' {\em Journal of the Acoustical Society of America}, vol.~77, pp.~806--812, 1985.

\bibitem{Duquennoy1999InfluenceON}
M.~Duquennoy, M.~Ouaftouh, M.~Ourak, and W.~J. Xu, ``Influence of natural and initial acoustoelastic coefficients on residual stress evaluation: Theory and experiment,'' {\em Journal of Applied Physics}, vol.~86, pp.~2490--2498, 1999.

\bibitem{HUANG2023102832}
C.-L. Huang, Y.~Wu, X.~He, M.~Dersch, X.~Zhu, and J.~S. Popovics, ``A review of non-destructive evaluation techniques for axial thermal stress and neutral temperature measurement in rail: Physical phenomena and performance assessment,'' {\em NDT \& E International}, vol.~137, p.~102832, 2023.

\bibitem{LIU2022108603}
E.~Liu, Y.~Liu, Y.~Chen, X.~Wang, H.~Ma, C.~Sun, and J.~Tan, ``Measurement method of bolt hole assembly stress based on the combination of ultrasonic longitudinal and transverse waves,'' {\em Applied Acoustics}, vol.~189, p.~108603, 2022.

\bibitem{WANG2022108500}
Y.~Wang, X.~Zhu, Y.~Gong, N.~Liu, Z.~Li, Z.~Long, and J.~Teng, ``Combination of transverse and longitudinal ultrasonic waves for plane stress measurement of steel plates,'' {\em Applied Acoustics}, vol.~188, p.~108500, 2022.

\bibitem{CAO2002171}
Y.~Cao, S.~Li, and L.~Petzold, ``Adjoint sensitivity analysis for differential-algebraic equations: algorithms and software,'' {\em Journal of Computational and Applied Mathematics}, vol.~149, no.~1, pp.~171--191, 2002.
\newblock Scientific and Engineering Computations for the 21st Century - Me thodologies and Applications Proceedings of the 15th Toyota Conference.

\bibitem{2022-031}
Z.~Hejun, L.~Qinya, and Y.~Jidong, ``Recent progress on full waveform inversion,'' {\em Reviews of Geophysics and Planetary Physics}, vol.~54, no.~3, pp.~287--317, 2023.

\bibitem{nolet1987seismic}
G.~Nolet, ``Seismic wave propagation and seismic tomography,'' in {\em Seismic tomography: With applications in global seismology and exploration geophysics}, pp.~1--23, Springer, 1987.

\bibitem{10.1190/1.3238367}
J.~Virieux and S.~Operto, ``{An overview of full-waveform inversion in exploration geophysics},'' {\em Geophysics}, vol.~74, pp.~WCC1--WCC26, 12 2009.

\bibitem{10.1111/j.1365-246X.2004.02453.x}
J.~Tromp, C.~Tape, and Q.~Liu, ``{Seismic tomography, adjoint methods, time reversal and banana-doughnut kernels},'' {\em Geophysical Journal International}, vol.~160, pp.~195--216, 01 2005.

\bibitem{10.3389/feart.2022.886920}
H.~Yang, L.-Y. Fu, B.-Y. Fu, and T.~M. Müller, ``Acoustoelastic fd simulation of elastic wave propagation in prestressed media,'' {\em Frontiers in Earth Science}, vol.~10, 2022.

\bibitem{egle1976measurement}
D.~Egle and D.~Bray, ``Measurement of acoustoelastic and third-order elastic constants of rail steel,'' {\em The journal of the Acoustical Society of America}, vol.~59, no.~S1, pp.~S32--S32, 1976.

\bibitem{radaj2012heat}
D.~Radaj, {\em Heat effects of welding: temperature field, residual stress, distortion}.
\newblock Springer Science \& Business Media, 2012.

\bibitem{leon1996residual}
T.~Leon-Salamanca and D.~Bray, ``Residual stress measurement in steel plates and welds using critically refracted longitudinal (lcr) waves,'' {\em Research in Nondestructive Evaluation}, vol.~7, pp.~169--184, 1996.

\bibitem{CUICongyue_246}
W.~Y. CUI~Congyue, ``A synthetic study on full seismic waveform inversion for one dimensional velocity structure,'' {\em Acta Scientiarum Naturalium Universitatis Pekinensis}, vol.~55, no.~2, pp.~246--252, 2019.

\bibitem{GIVOLI2021113810}
D.~Givoli, ``A tutorial on the adjoint method for inverse problems,'' {\em Computer Methods in Applied Mechanics and Engineering}, vol.~380, p.~113810, 2021.

\bibitem{Fichtner2011FullSW}
A.~Fichtner, ``Full seismic waveform modelling and inversion,'' 2011.

\bibitem{10.1785/0120060041}
Q.~Liu and J.~Tromp, ``{Finite-Frequency Kernels Based on Adjoint Methods},'' {\em Bulletin of the Seismological Society of America}, vol.~96, pp.~2383--2397, 12 2006.

\bibitem{j.1365-246X.2006.02978.x}
R.-E. Plessix, ``A review of the adjoint-state method for computing the gradient of a functional with geophysical applications,'' {\em Geophysical Journal International}, vol.~167, no.~2, pp.~495--503, 2006.

\bibitem{WANG2018117}
W.~Wang, C.~Xu, Y.~Zhang, Y.~Zhou, S.~Meng, and Y.~Deng, ``An improved ultrasonic method for plane stress measurement using critically refracted longitudinal waves,'' {\em NDT \& E International}, vol.~99, pp.~117--122, 2018.

\bibitem{LUO2023102765}
Z.~Luo, H.~Wang, S.~Zhang, F.~Li, and S.~Jin, ``A phased array ultrasonic-based enhanced strategy of critically refracted longitudinal (lcr) wave technique,'' {\em NDT \& E International}, vol.~133, p.~102765, 2023.

\bibitem{ILKERYELBAY201029}
H.~{Ilker Yelbay}, I.~Cam, and C.~{Hakan Gür}, ``Non-destructive determination of residual stress state in steel weldments by magnetic barkhausen noise technique,'' {\em NDT \& E International}, vol.~43, no.~1, pp.~29--33, 2010.

\bibitem{JAVADI2013628}
Y.~Javadi, M.~Akhlaghi, and M.~A. Najafabadi, ``Using finite element and ultrasonic method to evaluate welding longitudinal residual stress through the thickness in austenitic stainless steel plates,'' {\em Materials \& Design}, vol.~45, pp.~628--642, 2013.

\bibitem{lu2008ultrasonic}
H.~Lu, X.~Liu, J.~Yang, S.~Zhang, and H.~Fang, ``Ultrasonic stress evaluation on welded plates with lcr wave,'' {\em Science and Technology of Welding and Joining}, vol.~13, no.~1, pp.~70--74, 2008.

\bibitem{joseph2015evaluation}
A.~Joseph, P.~Palanichamy, and T.~Jayakumar, ``Evaluation of residual stresses in carbon steel weld joints by ultrasonic l cr wave technique,'' {\em Journal of Nondestructive Evaluation}, vol.~34, no.~1, p.~266, 2015.

\bibitem{Blaschke2017}
D.~N. Blaschke, ``Averaging of elastic constants for polycrystals,'' {\em Journal of Applied Physics}, vol.~122, no.~14, 2017.
\newblock Cited by: 19; All Open Access, Green Open Access.

\bibitem{baskaran2007simulation}
G.~Baskaran, C.~L. Rao, and K.~Balasubramaniam, ``Simulation of the tofd technique using the finite element method,'' {\em Insight-Non-Destructive Testing and Condition Monitoring}, vol.~49, no.~11, pp.~641--646, 2007.

\end{thebibliography}
\end{document}